\documentclass{article}
\PassOptionsToPackage{numbers, compress}{natbib}
\usepackage[preprint]{neurips_2022}
\usepackage{listings}
\usepackage{xcolor}

\definecolor{codegreen}{rgb}{0,0.6,0}
\definecolor{codegray}{rgb}{0.5,0.5,0.5}
\definecolor{codepurple}{rgb}{0.58,0,0.82}
\definecolor{backcolour}{rgb}{0.95,0.95,0.92}

\lstdefinestyle{mystyle}{
    backgroundcolor=\color{white},   
    commentstyle=\color{blue},
    keywordstyle=[1]\color{blue},
    keywordstyle=[2]\color{magenta},
    keywordstyle=[3]\color{blue},
    keywordstyle=[4]\color{codegray},
    numberstyle=\tiny\color{codegray},
    stringstyle=\color{codegreen},
    basicstyle=\ttfamily\scriptsize,
    breakatwhitespace=false,         
    breaklines=true,                 
    captionpos=b,                    
    keepspaces=true,                 
    numbers=left,                    
    numbersep=5pt,                  
    showspaces=false,                
    showstringspaces=false,
    showtabs=false,                  
    tabsize=2,
}
\lstdefinelanguage{mlir}{
    alsoletter={\%,\#,!,.,_},
    morekeywords={\%},
    keywords=[1]{func, return, attributes},
    keywords=[2]{tile, slice, sum, yield, atomic, spmd, range, all_sum},
    keywords=[4]{tensor, f32},
    showstringspaces=false,
	breaklines=true,
    breakatwhitespace=true,
    morestring=[b]",
    morecomment=[l]{//},
}
\usepackage[utf8]{inputenc} 
\usepackage[T1]{fontenc}    
\usepackage{hyperref}       
\usepackage{url}            
\usepackage{booktabs}       
\usepackage{amsfonts}       
\usepackage{nicefrac}       
\usepackage{microtype}      
\usepackage{xcolor}         

\usepackage[frozencache,cachedir=.]{minted}

\setminted{fontsize=\small}
\usepackage{graphicx}
\usepackage{wrapfig}

\usepackage{subcaption}

\title{Automatic Discovery of Composite SPMD Partitioning Strategies in PartIR}

\author{%
  Sami Alabed ~\thanks{Correspondence: sa894@cam.ac.uk. Work done during an internship at DeepMind.} \qquad
  Dominik Grewe \qquad
  Juliana Franco \qquad
  Bart Chrzaszcz  \AND
  Tom Natan \qquad
  Tamara Norman \qquad
  Norman A. Rink \AND
  Dimitrios Vytiniotis \qquad
  Michael Schaarschmidt \qquad \\
  \\ 
  DeepMind \\
}

\begin{document}

\maketitle
\begin{abstract}
Large neural network models are commonly trained through a combination of advanced parallelism strategies in a single program, multiple data (SPMD) paradigm. 
For example, training large transformer models requires combining data, model, and pipeline partitioning; and optimizer sharding techniques. 
However, identifying efficient combinations for many model architectures and accelerator systems requires significant manual analysis. 
In this work, we present an automatic partitioner that identifies these combinations through a goal-oriented search. 
Our key findings are that a Monte Carlo Tree Search-based partitioner leveraging partition-specific compiler analysis directly into the search and guided goals matches expert-level strategies for various models.

\end{abstract}

\section{Introduction}
The rapid rise of large neural networks with significant memory and compute requirements have made partitioning strategies critical for enabling their training and inference. For example, large language models such as GPT-3 \cite{gpt3_2020}, Gopher \cite{gopher2021}, PaLM \cite{palm2022} or Chinchilla \cite{chinchilla2022} have relied on various combinations of Megatron-style layer sharding \cite{megatron2019, large_megatron_21}, batch/data parallelism, pipeline parallelism \cite{pipedream, pipemare21, gpipe}, or ZeRO sharding/offloading \cite{zero2019, zero_offload2021}.

Model sizes across applications have been outpacing device memory and FLOPS growth: Google TPU \cite{DBLP:journals/corr/JouppiYPPABBBBB17, jouppi+:lessons} v2 reports 46 TFLOPS / 16 GB of HBM per chip, while v4 reports 275 TFLOPS / 32 GB\footnote{\url{https://cloud.google.com/tpu/docs/system-architecture-tpu-vm}} over a period where model parameters -- and consequently FLOPS requirements -- increased by approximately 4 orders of magnitude.
The use of sophisticated parallelism strategies is, therefore, unavoidable. Despite progress in the development of partitioning APIs (e.g. GSPMD-style \cite{gspmd2021, gshard} partitioning annotations exposed in front-end APIs by JAX \cite{jax2018github}) or parallelism frameworks such as DeepSpeed \cite{deepspeed_2021}, customising new research models to hardware configurations often still requires expert analysis to account for model-specific compute and communication patterns.
Recent work has proposed automated parallelism frameworks specifically targeting large neural network training. For example, Alpa \cite{alpa2022} defines a hierarchical space of intra- and inter-operator parallelism and solves sub-problems through integer linear programming (ILP). 
Unity \cite{unity2022} as an extension of FlexFlow \cite{flexflow2019} and TASO \cite{taso2019} jointly optimises parallelism and algebraic transformations through hierarchical search. 
There exists work on applying RL for the partitioning problem~\cite{wang+:automap}, recently combining RL with a constraint solver and achieving good generalization properties~\cite{xie:transferrable}.

We approach automated parallelism with a different philosophy. 
Our automated SPMD partitioner is closely integrated with our PartIR compiler stack. 
PartIR \cite{partir2020} provides our partitioner with the ability to validate partitioning decisions, expose a host of static analyses about the module, automatically propagate sharding
decisions across a module, and produce analytical cost model
estimates. The latter is obtained using a simulator of a lowered and communication-optimized partitioned module at the compiler level. By contrast, ILP-based encodings~\cite{alpa2022}
attempt to address scalability issues in large graphs by e.g.
heuristically grouping operators to prune the search space effectively -- furthermore, there may be communication
optimizations post-ILP that cannot be captured when deciding the parallel plan. FlexFlow~\cite{flexflow2019} addresses scalability by making the simulator in their MCMC search incremental and relies on pre-defined layers/block-operators (e.g. Multi-Head Attention). 
On the other hand, our approach can partition {\em any} array operation of a low-level tensor IR (XLA) that high-level NN libraries compile. As we show, 
our tight compiler integration allows our search to be data and size efficient and run routinely as part of our user (typically researchers) iterative workflows.

In this work, we extend our automated partitioner, Automap \cite{automap}, to support partitioning a model across multi-dimensional device topology and discover expert-level composite SPMD partitioning strategies.
Our contribution is a design of a goal-oriented Monte Carlo tree search (MCTS) \cite{Browne12asurvey} that decomposes the problem into smaller sub-problems.
Furthermore, we incorporate a partitioning-specific compiler analysis into the MCTS to reduce both the nodes and edges of the tree, improving the search's robustness.
We show that our partitioner discovers composite expert-level SPMD strategies for common models, such as Transformers and GraphNets.
Moreover, it produces significantly better than human-written strategies for models
as UNet, for which no known strategy is available.

\section{Background}
\subsection{Logical Meshes and Composite Strategies}
\begin{figure}[htbp]
  \begin{minipage}[c]{0.3\textwidth}
    \includegraphics[scale=0.3]{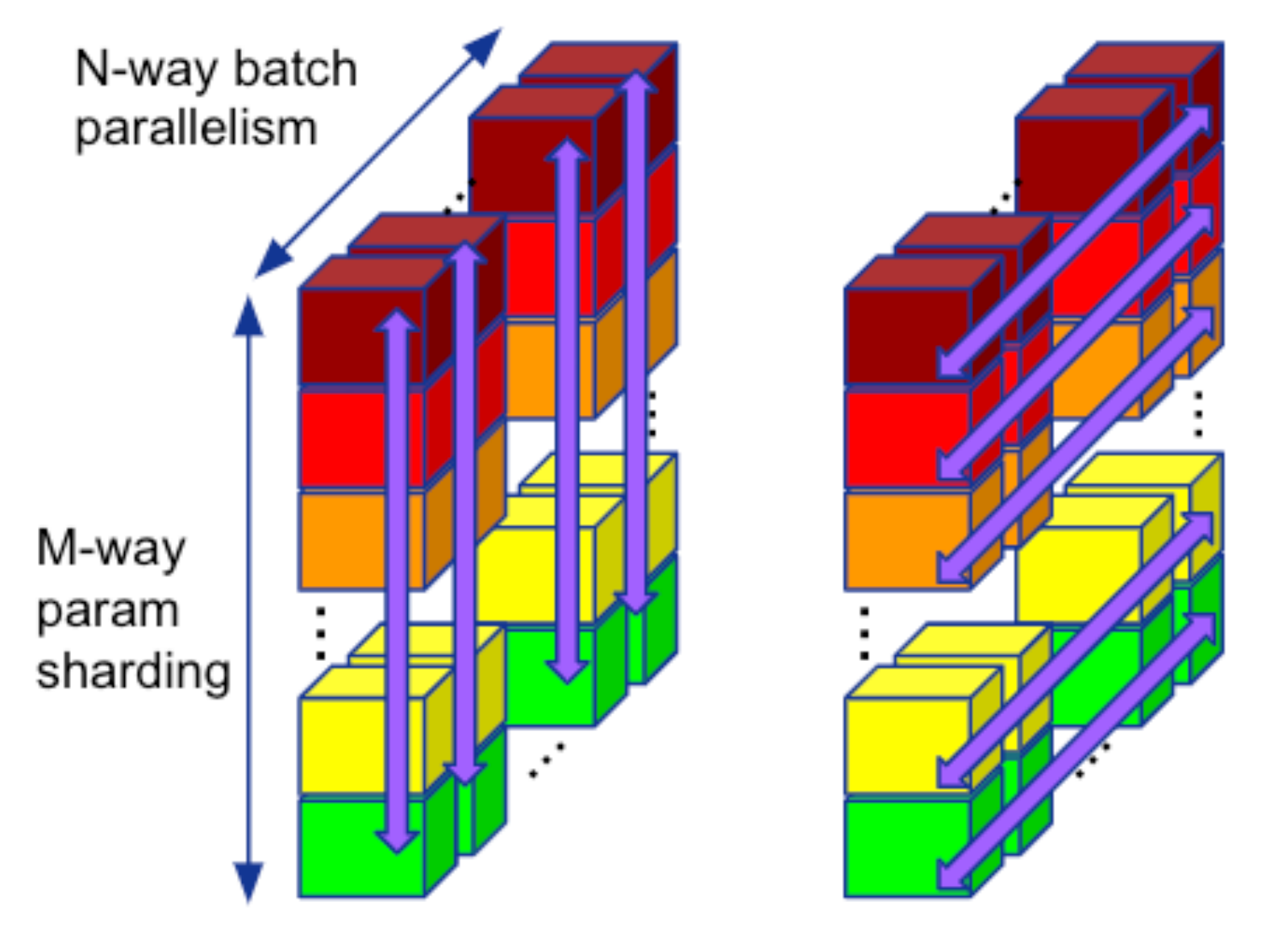}
  \end{minipage}
  \hfill
  \begin{minipage}[c]{0.5\textwidth}
    \caption{\small The composite strategy of batch and model parallelism over mesh 
$\{\texttt{batch}{:}N, \texttt{model}{:}M\}$. We also show communication patterns that may emerge; on the left, possible communication along the {\tt model} axis (e.g., Megatron activation reductions), and on 
the right, communication along the {\tt batch} axis (e.g., gradient reductions). The color
coding denotes the unique parameter shards that each device holds; e.g., all devices along the {\tt batch}
axis holds the same shard of parameters.}\label{fig:logical-partitioning}
  \end{minipage}
\end{figure}

Leveraging composite partitioning techniques has enabled training of recent large models \cite{gpt3_2020, gopher2021, chinchilla2022}. The main idea is to structure the available accelerator devices into an n-dimensional logical {\em mesh} (which will typically, but not necessarily, correspond to the physical
communication topology). For instance, we may view 32 TPU devices as a 4x8 mesh or a system of 2 GPU servers with 16 GPUs each as a 2x16 mesh. Once we are given such a logical mesh of
available devices, e.g., with a 2D mesh, a conventional strategy would be to do batch parallelism over one axis
and parameter sharding (model parallelism) over the other. Figure~\ref{fig:logical-partitioning} graphically
depicts this strategy. ZeRO-style sharding of the optimizer~\cite{zero2019} (on top of batch parallelism,
and possibly also parameter sharding) is simply a different stage that shards the {\em optimizer parameters} along the axis used for {\em batch parallelism}.

Conceptually, each stage of a composite strategy, like the above, optimizes for specific objectives.
For example, batch parallelism and parameter sharding typically improve runtime (while also reducing memory requirements); 
whereas ZeRO-style optimizer sharding aims mainly towards improving memory (but may improve runtime too, as the typically memory-bound vector operations of the optimizer update are sharded, as
was already observed in precursor work~\cite{weight-update-sharding}). ZeRO ``stage-2'' optimizer sharding may not increase communication cost 
since it replaces AllReduce operations with pairs of parameter AllGather and gradient ReduceScatter. ZeRO ``stage-3'' aims to further improve memory
by introducing separate parameter AllGather for the forward and the backward pass of a model; hence may slightly increase the runtime in favor of keeping smaller
tensors live across forward and backward computation.
Note that, in our setting, the logical mesh will be given ahead of time by the user; the partitioner's task is to discover composite strategies like those described above based on user-provided objectives.

\subsection{PartIR rewriting for SPMD partitioning}
The PartIR compiler infrastructure accepts a user-defined ML model in the form of an
MLIR~\cite{mlir2020} module, a user-provided mesh, and exposes rewrite actions on that 
module to express forms of shardings. 
The key idea behind PartIR rewriting is to express
parallelism of array operations, like {\tt z = matmul(x, w)}, as parallel 
{\em tiling} or {\em reduction} loops, specifically tagged with the axis over which these parallel loops range.
e.g:
\begin{center}
\begin{tabular}{c|c}
\begin{lstlisting}[language=mlir]
// Tiling loop derived from sharding x
// along dimension 0 on "batch" axis.
%z = tile 0 "batch" (%r : range<N>) {
  %xs = slice 0 %x[%r]
  %zs = matmul(%xs, %w)
  yield %zs
}
\end{lstlisting} & 
\begin{lstlisting}[language=mlir]
// Reduction loop derived from sharding
// both inputs along a contracting 
// dimension on "model" axis.
%z = sum "model" (%r : range<M>) {
  %xs = slice 1 %x[%r]
  %ws = slice 0 w[%r]
  %zs = matmul(%xs, %ws)
  yield %zs
}
\end{lstlisting}
\end{tabular}
\end{center}
Compiler-exposed sharding actions,
of the form ${\tt partition}(parameter\_id,\;dimension,\;axis)$, introduce and 
{\em propagate} such tiling loops across the MLIR module, based on powerful propagation rules that
depend on the semantics of the array operations in hand. 

\paragraph{Search with PartIR actions}
Automap~\cite{automap} applies such actions to the module, guided by a \emph{worklist} consisting of the module arguments (e.g., model parameters, optimizer state and the data tensors passed to a training step function). 
If the compiler detects that the sharding of an argument can be propagated to another argument, then both are removed from the worklist and no longer considered for further sharding actions (See Figure~\ref{fig:worklist} for an example.)

\begin{figure*}
\begin{center}
\begin{tabular}{c|c}
\begin{lstlisting}[language=mlir]
// worklist = {%arg0, %arg1, %arg2}
func f(%arg0, %arg1, %arg2) {
  %0 = add(%arg0, %arg1)
  ...
}
\end{lstlisting} & 
\begin{lstlisting}[language=mlir]
// worklist = {%arg2}
func f(%arg0, %arg1, %arg2) {
  %0 = tile 0 "x" (%r : range<N>) {
    %slice0 = slice 0 %arg0[%r]
    %slice1 = slice 0 %arg1[%r]
    %a = add(%slice0, %slice1)
    yield %a
  }
  ...
}
\end{lstlisting}
\end{tabular}
\end{center}
\caption{Sharding argument {\tt \%arg0} along axis {\tt batch} in the left module causes {\tt \%arg1} to also become sharded; leaving us with a worklist of just {\tt \%arg2} and the rewritten module on the right.}\label{fig:worklist}
\end{figure*}

As a result, a very small number of sharding actions can lead to high-quality partitions (e.g., low communication overhead) within a reasonable search budget. 
This can be amplified by further compressing the worklist with user-provided groupings of parameters that should be equi-sharded (e.g. all transformer blocks across a large transformer model). 
The overall search is driven by a cost model, typically simulated runtime cost and peak per-device memory obtained statically.

\section{Multi-axis search}
In this work, we extend Automap~\cite{automap} to support multi-axis automatic partitioning, a capability that our 
PartIR stack already supports through {\em nesting}
tiling and reduction loops, as well as multi-axis loop 
propagation.
However, multi-axis partitioning increases the search complexity exponentially for every axis.
The search must consider the dependency between partitioning decisions and the argument for partition, axis, and dimension. Compiler static analysis together with Automap's~\cite{automap} worklist mechanism described above helps us prune the partitioning decision branching factor.
This section outlines the major additions needed to reduce the complexity of the problem and reach fast expert-level sharding on several representative models. 

\subsection{Monte Carlo Tree Search for SPMD compilation}
The automated partitioner is implemented as a Monte Carlo Tree Search (MCTS) \cite{Browne12asurvey} with upper confidence bound for trees (UCT). The MCTS starts with no prior knowledge about the model's partition, and by exchanging with PartIR partitioning actions for performance estimates, the MCTS builds a policy that allows it to explore the partitioning space efficiently.
Each action in the MCTS translates to a one-hot encoding of a partitioning decision (argument, dimension and mesh axis). 
The reward is the estimated performance calculated by PartIR and observed at the end of an episode.

The main users of our partitioner are researchers who partition their models during iterative research workflows. 
In the next section, we discuss how the search can discover effective solutions quickly to enable fast research iterations.

\subsubsection{Domain specific enhancements over Automap}
As in Automap \cite{automap}, we leverage user equi-sharding annotations to partition repeated layers all at once, as if partitioning a single layer,
reducing the search space further. Moreover, we introduce several domain-specific enhancements into the MCTS to improve its efficiency and reduce the complexity of the problem in preparation for the multi-axis search.

\paragraph{Compressed state representation}
In the original Automap work, \cite{automap}, the sequences of partitioning decisions identified the module's state in the MCTS. 
However, PartIR infers and propagates partitioning decisions of a single argument to other values across the module.
For example, in Figure~\ref{fig:worklist}, we can arrive to the module and worklist on the right by {\em either} partitioning {\tt \%arg0} or {\tt \%arg1}. 
This leads to an interesting observation; there could be several
permutations of decisions within the same axis that lead to a similar final partitioned model and increase the branching factor of the search needlessly.
PartIR provides static analysis of the state
of each variable of the model (e.g., on which axes and dimensions it has already
been sharded). 
We leverage this analysis to identify how partitioning actions impact the module's worklist. 
Such impact is then used to identify the state of the module. 
This enhancement compresses the state-space of the MCTS, as there are fewer {\em truly} unique states - akin to reducing the number of nodes in the tree, allowing the search to explore previously infeasible regions to reach.

\paragraph{Action grouping}
Building on the previous finding, various partitioning decisions (actions) can lead to the same state after propagation.
This knowledge would save the search from having to take action and find that out - akin to collapsing the out edge from the node, and effectively further reducing the branching factor of the tree \cite{actiongrouping}. To achieve this, upon MCTS backpropagation, we assign the same reward to any actions that lead to the same worklist sharding state. As a result, the MCTS expansion policy does not have two consider separate actions that lead to the same worklist state, effectively pruning the MCTS branching factor.

\subsubsection{Meta-controller}
Domain-specific enhancements reduce the search branching factor; however, the question of scaling to handle the multi-axis complexity remains. 
This work proposes using a meta-controller to focus the search effort on a single axis and a given objective.
The meta-controller decomposes the global objective into a hierarchy of goals of the form $(axis, objective)$. The order of goals influences the overall partitioning objective and reflects the expert's preferences.
The meta-controller allocates a fixed budget to each goal and guides the search to focus on one of them.
Once a goal exhausts its allocated budget, the meta-controller analyses the best-found partitioning strategy and replaces the model with the partitioned model if it finds it would improve on the goal's objective. 
After that, the search focus on the next goal.
This design enables the search to discover composite strategies by independently searching for partitioning strategies on each axis, only considering a single objective.
Exposing these scripted goals to the users allows them to encode bespoke requirements such as finer control over the trade-off between runtime and memory.

\subsubsection{MCTS enhancements evaluation}
This section evaluates how the enhancements described above enable Automap to work efficiently in a multi-axis setting.
Transformers are well understood and studied in the DNN scaling literature \cite{scaling_laws_2020, chinchilla2022} and existing frameworks, such as DeepSpeed \cite{deepspeed_2021} Megatron-LM \cite{megatron2019, large_megatron_21}, implement well-known partitioning strategies for their partition (e.g. Megatron \cite{megatron2019, large_megatron_21} and ZeRO \cite{zero2019}). 
A success metric for our work is to automatically discover Batch and Megatron model parallelism (BP and MP, respectively), together with ZeRO-3 partition, given a Transformer model. 

\begin{figure}[htbp]
\begin{subfigure}{0.45\textwidth}
 \includegraphics[width=\textwidth]{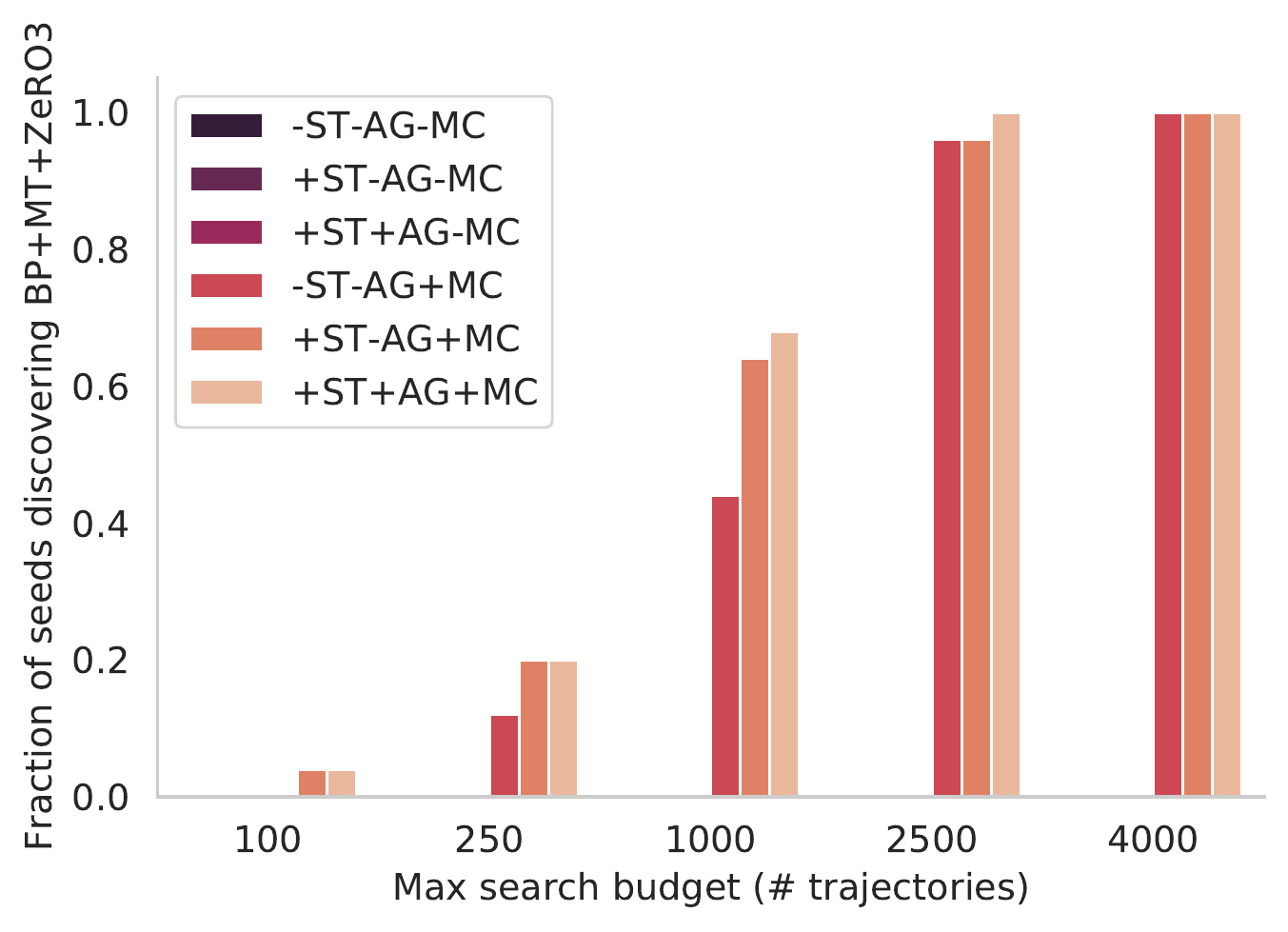}
\caption{\small The odds that search discovers the composite sharding strategy "BP+MT+ZeRO3" in 25 seeds. \label{fig:enhancement_eval}}
\end{subfigure}
\hfill
\begin{subfigure}{0.45\textwidth}
\vspace{4.8mm}
 \includegraphics[width=\textwidth]{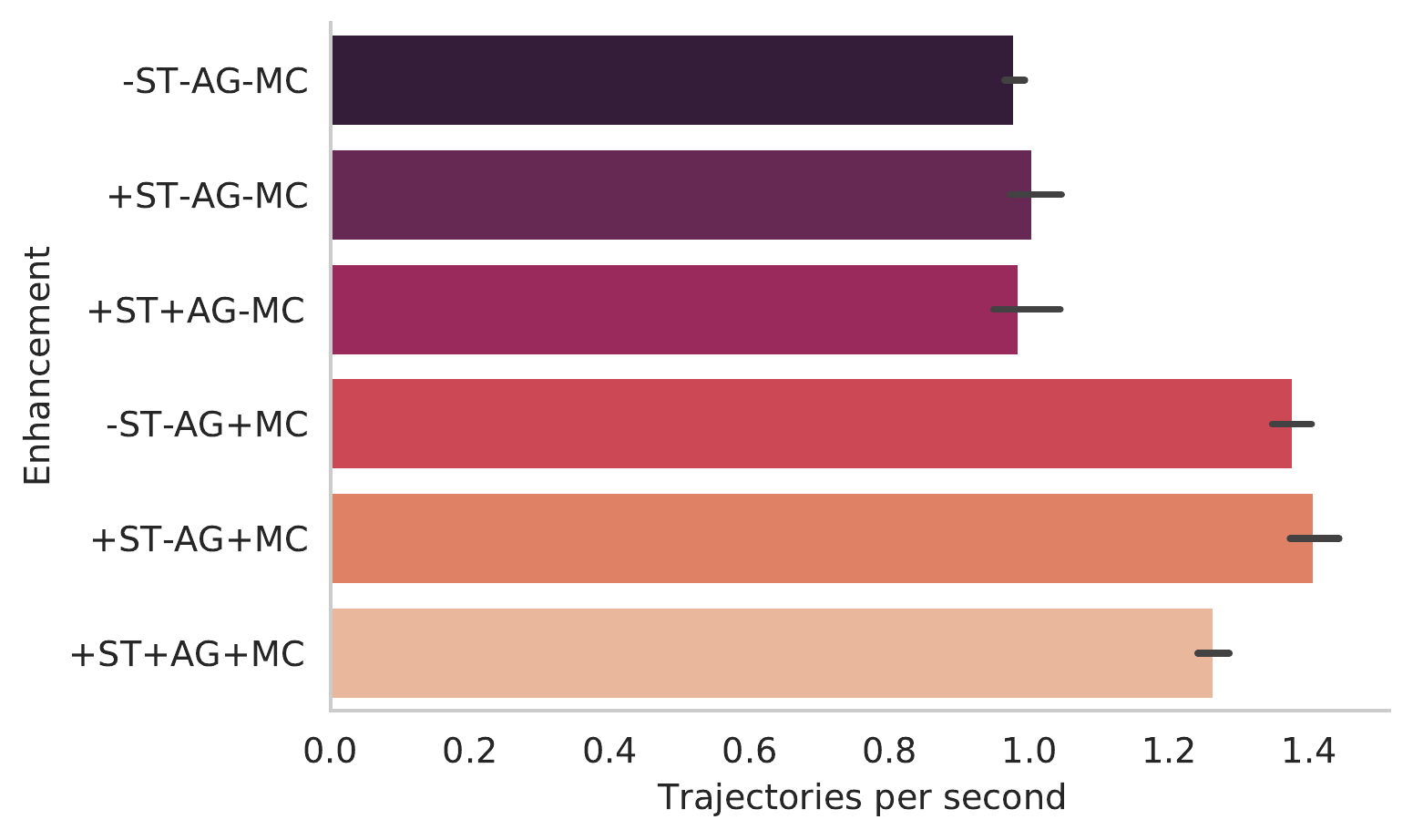}
\caption{\small The enhancements impact on the search throughput. 
We report the average for $25$ different seeds. Higher throughput is better. \label{fig:enhancement_runtime}}
\end{subfigure}
\caption{MCTS enhancement that enables multi-axis search. 
ST: the state representation enhancement, AG: action grouping, MC: meta-controller. $+$ indicates the enhancement is enabled. 
Note that it is impossible to do action grouping without partitioned arguments as the state.}
\end{figure}

Figure \ref{fig:enhancement_eval} shows the benefit of using the MCTS enhancements in improving the odds of discovering the composite sharding strategy of BP+MT+ZeRO3.
In all cases, only with the aid of the meta-controller was the search able to discover the composite strategy.
Enabling each enhancement improved the odds of discovering the strategy, especially in constrained budget scenarios.

The search runs as part of the researcher development workflow and needs to be quick. In production, our search budget is, on average, $\approx1000-2000$ trajectories.
As the search is constrained for trajectories, the enhancements must not be negatively impacting its throughput. Figure \ref{fig:enhancement_runtime} evaluates their impact. 
Using the meta-controller improves the search throughput significantly as it allows the rewrite engine to skip checking the validity of partitioning actions on the inactive axis as the meta-controller ensures the search is acting only on a single axis.  
The small reduction in throughput because of the state and action grouping enhancement is mitigated by the fact that they aid in discovering complex solutions sooner and are more robust, as shown in Figure~\ref{fig:enhancement_eval}.

Next, we investigate how the meta-controller utilizes its budget and allocate a large budget ($4000$ trajectories) to it.
Figure \ref{fig:discovering_expert} shows the trajectories needed to achieve each expert-level partitioning strategy.
First, the meta-controller divides the maximum budget evenly between all the goals, which has the side-effect that the search spends too many cycles on one goal; the search discovered BP's strategy in 50 trajectories, and it continued exploring until it exhausted its budget.
Similarly, the MT goal was achieved in the 250 trajectories. After that, however, it had to wait for the first goal to exhaust its budget before continuing.
The shaded bar shows the opportunity to improve the search ability to finish quicker if it utilizes its budget smarter. 
We leave this as a discussion point on removing the need for budget allocation or providing a smarter budget estimate for each axis.
Another opportunity for future work is the ability to discover these high-level goals automatically from data.
\begin{minipage}[htbp]{0.45\textwidth}
\centering
\begin{figure}[H]
 \includegraphics[width=1.\textwidth]{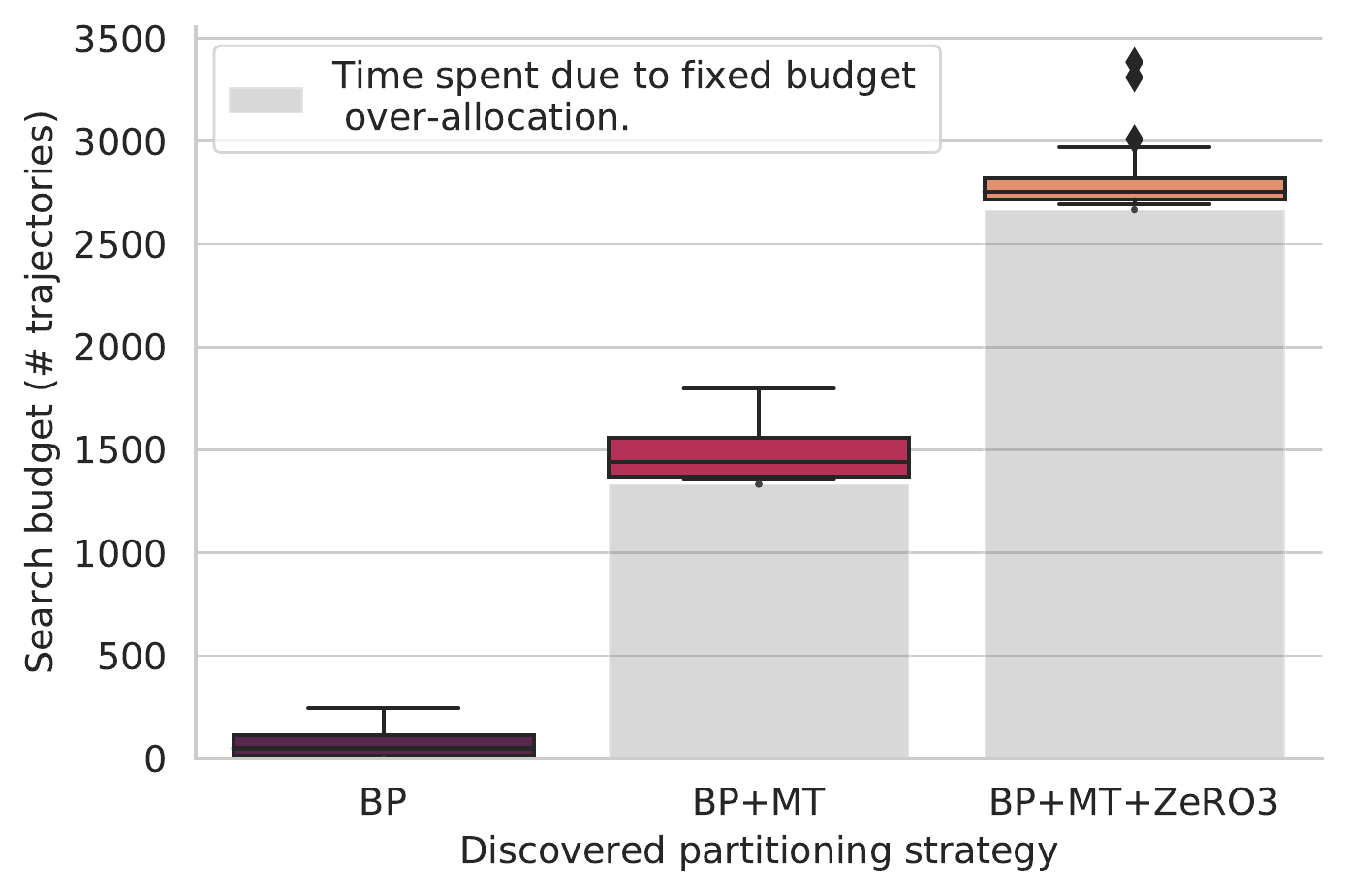}
\caption{\small The trajectories needed to discover expert partitioning strategies automatically. 
Validated by analyzing the collectives reported in Table \ref{tab:collectives}. \label{fig:discovering_expert}}
\end{figure}
\end{minipage}
\hspace*{0.75cm}     
\begin{minipage}[c]{0.45\textwidth}
\centering
\begin{table}[H]
\scriptsize
\caption{\small The number of SPMD collectives of the discovered partitioning strategies that correspond to manual expert partitioning. \label{tab:collectives}\\}
  \begin{tabular}{l p{4.5em} p{4.5em} p{4.5em}}
    \toprule
    Strategy & AllGather & AllReduce & ReduceScatter \\
    \midrule
    BP & $0$ & $387$ & $0$ \\
    BP+MT & $0$ & $483$ & $0$ \\
    BP+MT+ZeRO3 &  $578$ & $97$ & $386$ \\
    \bottomrule
  \end{tabular}
\end{table}
\end{minipage}

\section{Evaluation}
\textbf{Case-studies.} 
We evaluated our multi-axis search on three different model architectures:
\begin{itemize}
    \item A 24-layer transformer model similar to GPT-3 \cite{gpt3_2020} with input batch size of $16$, $d_{head}=96, n_{head}=24, d_{model}=1024$. 
    Training this model has an estimated 280GB memory requirement making it infeasible to fit on any single device.  
    \item A Graph Network Simulator (GNS) \cite{sanchez2020learning} model used in molecular structure prediction \cite{godwin2022simple}, configured with 24 layers of message passing, 1024 MLP hidden sizes as well as 512 latent sizes, with up to 10 molecule graphs per input batch (dynamically batched using jraph \cite{jraph2020github}).
    \item A UNet of the reverse process in a diffusion model \cite{unet_diffusion}, with a $32$ batch size. It uses $9$ residual blocks for the down-sampling convolutions and up-sampling one, with two residual blocks in-between and an attention layer at $16\times8\times4$ resolution with $n_{heads}=16$.
\end{itemize}

\textbf{Hardware.} The transformer experiments were conducted on 32 TPU v3 cores laid out in an 8x4 2D mesh. The UNet and GNS models are smaller and conducted on 8 TPU v3 cores in a 2x4 2D mesh. The hardware runtime was averaged over ten measurements following one warmup step.

\textbf{Meta-controller settings.}
A key design decision for the meta-controller is its ability to translate user requirements to meaningful search targets. At a high level, users want programs to fit into device memory and run as fast as possible. Here, we experiment with how different manually specified goals help achieve partitioning objectives. Eventually, we expect the meta-controller to handle the sequencing of objectives automatically. We tested four predefined goals:
\begin{itemize}
    \item RT\_MEM\_ALL: Improve runtime on each axis and then reduce the memory on each axis. 
    \item RT1\_RT2\_MEM1: Improve runtime on each axis and then aim to reduce the memory on the batch axis.
    This goal is based on the same rationale behind a ZeRO3-like partitioning.
    \item RT1\_RT2\_MEM2: Similar to the previous but reduces memory usage on the model axis. 
    \item RT\_MP\_ALL: Improve runtime on each axis while applying a progressively heavier memory penalty if the model does not fit in memory.
    \item No controller: No meta-controller, the search runs with a cost function that estimates runtime and a memory penalty term. The penalty term applies a linear cost based on exceeding the available memory to discourage partitions that do not fit.
\end{itemize}
As a simplification, we allowed the search a budget of $4000$ trajectories with our compiler (but no actual hardware compilation), and the search results were compiled and tested on hardware.

\begin{figure}[htbp]
\begin{subtable}{0.45\textwidth}
\footnotesize
\caption{\small The reference estimated initial non-partitioned models performance. \label{tab:initial_model_perf}\\}
  \centering
  \begin{tabular}{lll}
    \toprule
    Mode & Memory (GB) & Runtime (ms) \\
    \midrule
    Transformer24 & 284.0 & 8740.0 \\
    GNS & 12.592 & 521.48 \\
    UNet &  8.761 & 84.9 \\
    \bottomrule
  \end{tabular}
\end{subtable}
\hfill
\begin{subfigure}{0.45\textwidth}
\centering
\includegraphics[width=\textwidth]{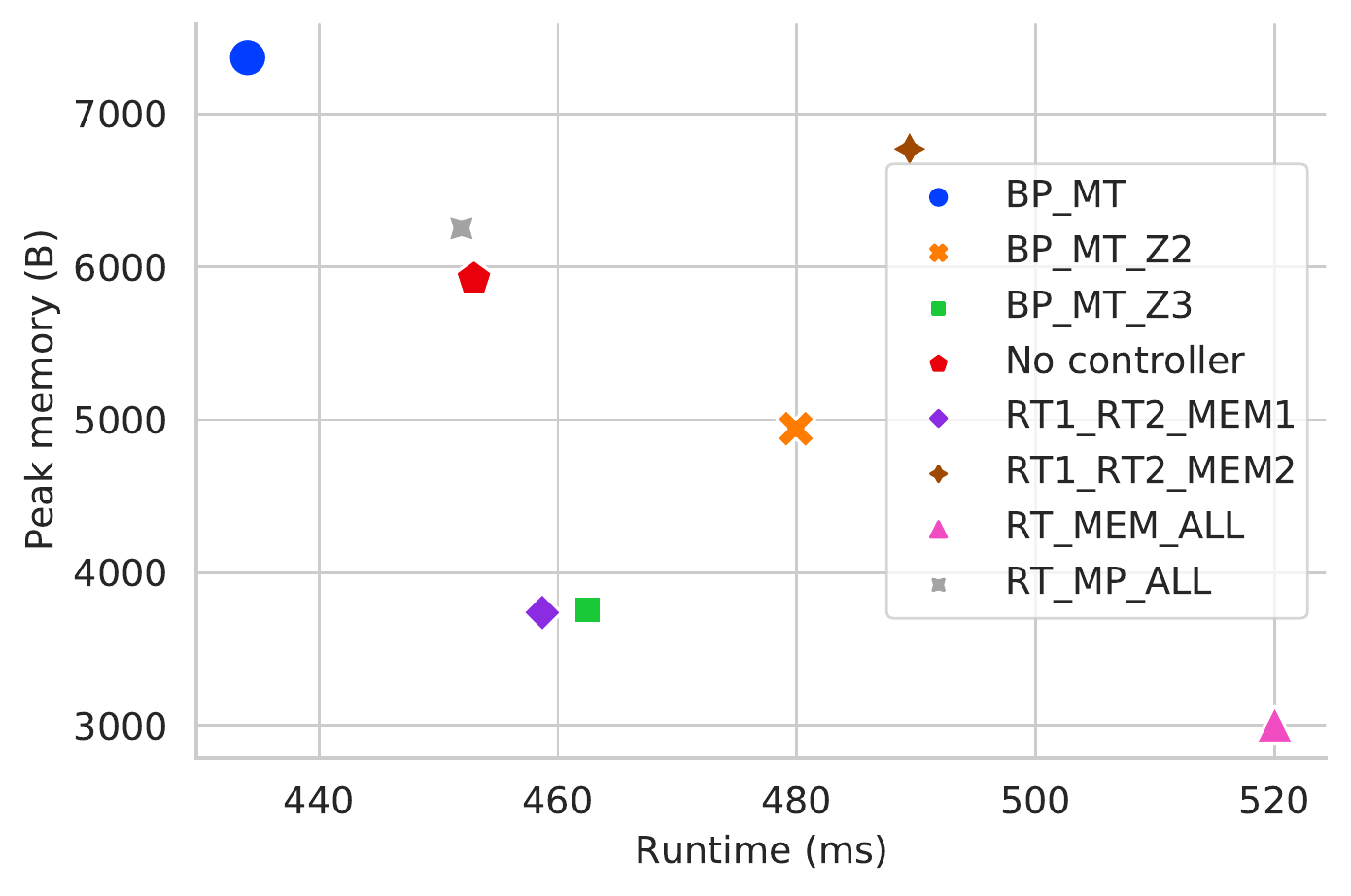}
\caption{\small A 24-layer transformer model. BP runs out of memory and is not in the figure. \label{fig:transformers_hw}}
\end{subfigure}
\vfill
\begin{subfigure}{0.45\textwidth}
 \includegraphics[width=\textwidth]{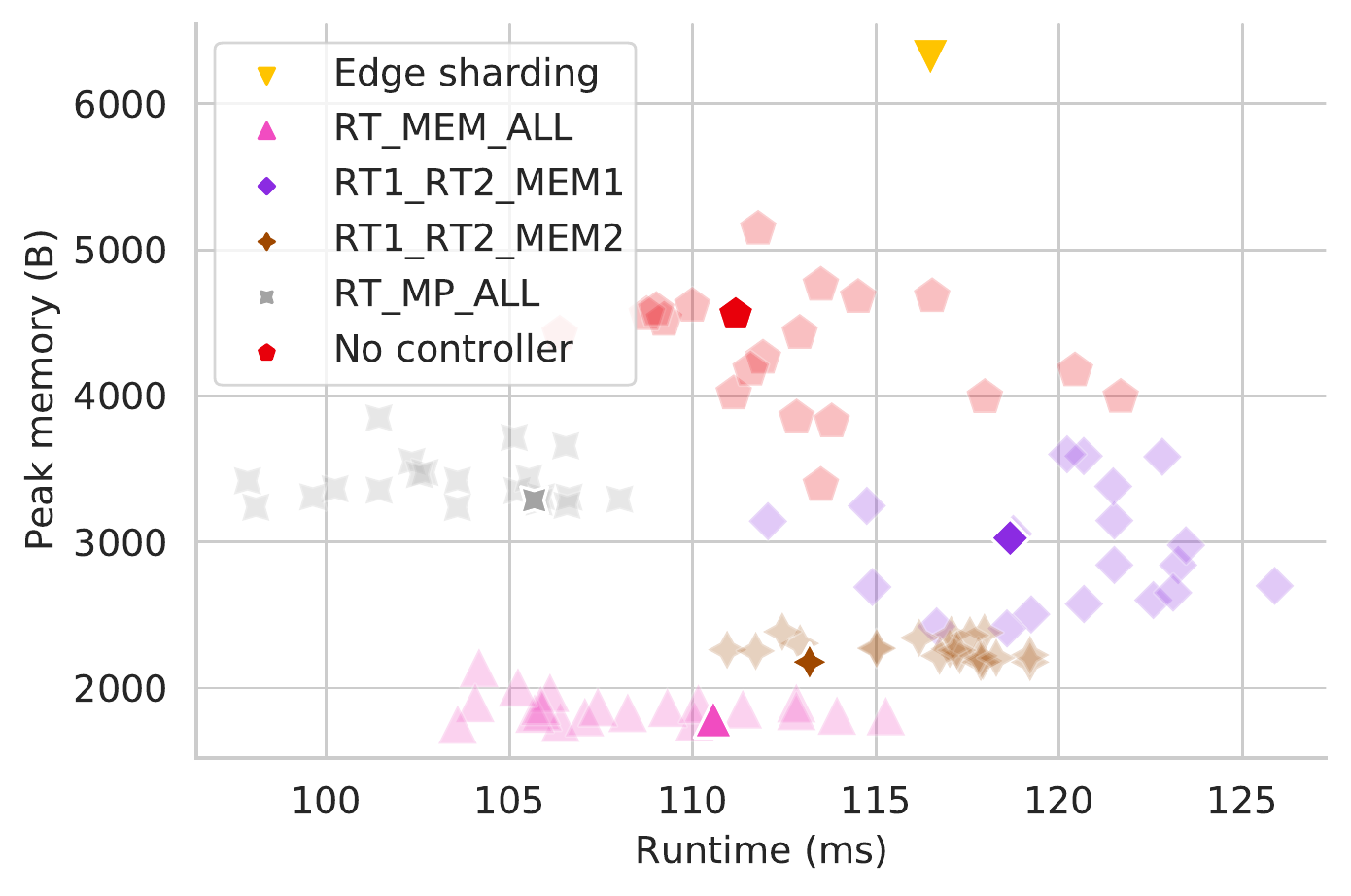}
\caption{\small A GNS model. \label{fig:gns_hw}}
\end{subfigure}
\hfill
\begin{subfigure}{0.45\textwidth}
 \includegraphics[width=\textwidth]{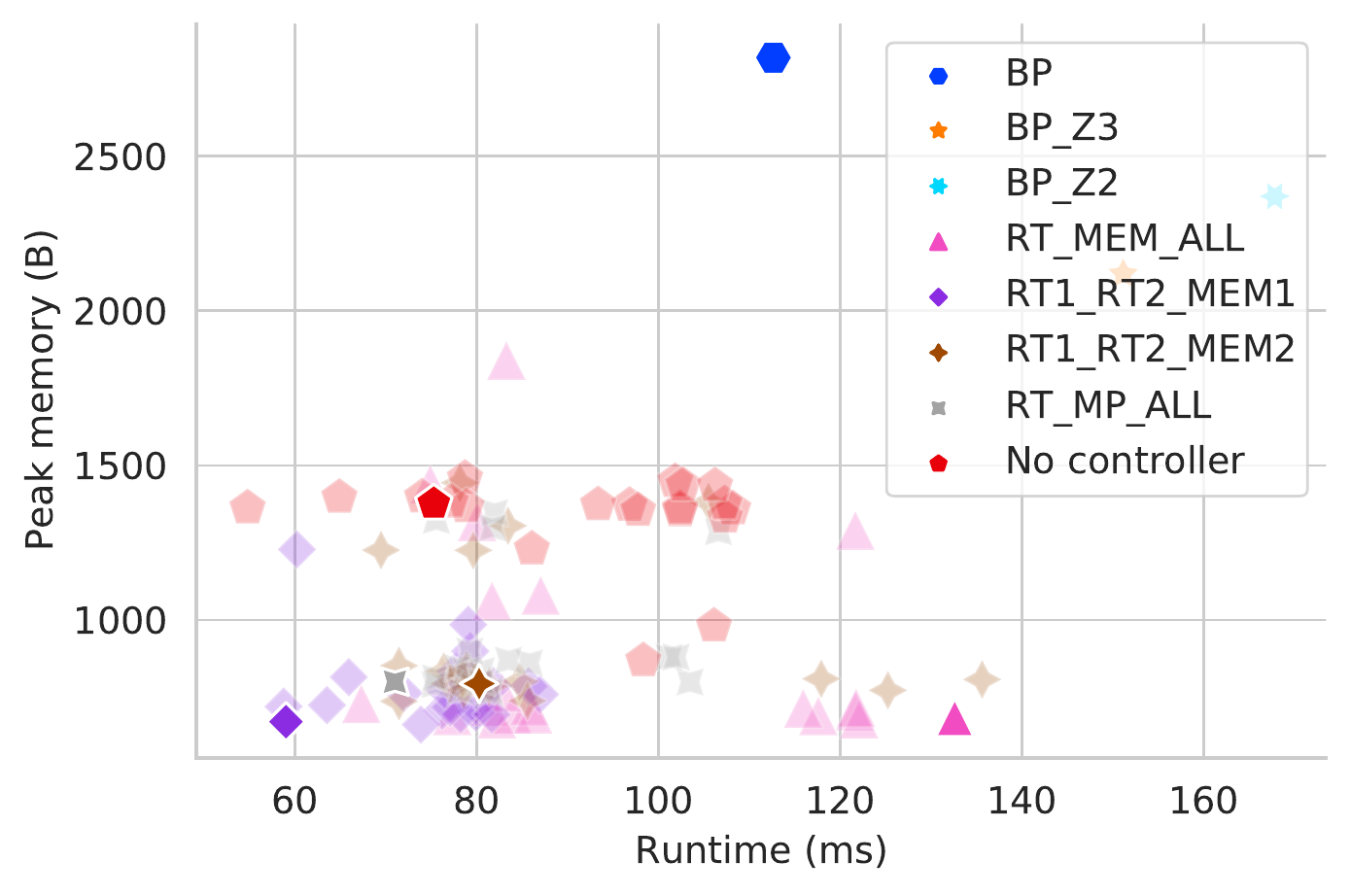}
\caption{\small A UNet like-model. \label{fig:unet_hw}}
\end{subfigure}
\caption{A runtime and memory performance for the models after partitioning using different partitioning strategies.
The lower values are better. Shaded points are from different search seeds, with the median search result in a bolder color.}
\end{figure}

In the remainder of this section, we evaluate the different meta-controller strategies and compare them to the known expert partitioning baselines.
Figure \ref{fig:transformers_hw} shows the runtime-memory plot once the model has been compiled and run on hardware. 
There is an obvious trade-off between the model's runtime and peak memory, with the leftmost and bottom-most being the best points of that trade-off.
These points drive the user requirement in designing the goals for the meta-controller.
For example, a user who needs a ZeRO3-like partitioning strategy that provides memory reduction while maintaining a high training throughput RT1\_RT2\_MEM1 is an ideal strategy.
However, a user constrained for memory would opt for a strategy focused on memory reduction. 
For example, in our initial experiments, we ran the same model with a 2x2 TPU topology (which has 16 devices). All strategies failed to fit on-device memory (even ZeRO3 could only get the model down to 17GB > 16GB available on TPU v3); all strategies except for the RT\_MEM\_ALL that was able to bring it down to 11GB by focusing on aggressive memory sharding, and still achieved an average of 1-second runtime (compared to the initial model estimate of 8 seconds).

To further validate the search performance, we evaluated it on other models of very different architecture: UNet and GNS.
UNet does not have a known best expert partitioning strategy. Therefore, we used naive batch parallelism with ZeRO3 as a baseline. For GNS, we use edge-sharding~\cite{edge_sharding} as a baseline. Edge-sharding shards the edges into sub-graphs on each axis, such that the computation is run concurrently. In particular, the original single-axis Automap has been used successfully to partition GNS-style models \cite{zaidi22} via edge sharding.

We can see in Figure \ref{fig:gns_hw} that GNS strategies focused on aggressively reducing the memory or runtime have a significant advantage over edge-sharding.
This illustrates the existence of non-trivial partitioning strategies that could partition the graph into smaller sub-graphs to run concurrently, and each device only needs to store a smaller subset of the parameter and optimizer states. Similarly, in Figure \ref{fig:unet_hw}, UNet exhibits an interesting point where there is a better policy than all other strategies. 
This evaluation demonstrates that there are models, which are not being actively researched in the scaling literature, that have the potential to train twice as fast with half the resources, even when compared to strategies that are thought to be very effective (batch parallelism and ZeRO3).  

\textbf{Action sequence length.} Using PartIR compiler stack and its powerful sharding propagation technique and grouping of arguments, the depth of the decision tree is shortened. 
To discover the BP+MT+ZeRO3 transformer sharding, the agent had to perform $13$ actions, in contrast, to discover our unique UNet solution, the search had to perform $30 \approx 40$ actions. 
Deciding the ideal depth of the tree leads to faster convergence as the search does not have to go deeper than it should and allows us to find unique solutions for models that require higher than average sequence of sharding decisions.
This decision can either be learned or integrated through the powerful compiler analyzer tool chain.

\section{Discussion and outlook} 
This work explores the idea of incorporating several optimization objectives (goals) to discover composite partitioning strategies and proposes to improve MCTS efficiency through compiler analysis. 
There are several simplifications made in this work that are worth revisiting. 
Budget estimation for each goal is non-trivial, but further compiler analysis to approximate the cardinality of the independent partitioning actions can be used to bound both episode lengths and budget.
Moreover, balancing local and global optimization goals is challenging and may require re-entrant search \cite{xtat2021}. Finally, searching over heterogeneous SPMD meshes or temporal/pipeline strategies has not been explored yet but opens additional opportunities for meta-controller improvements. We also point out that leveraging compiler analysis in search is not free, and cost increases with model scale. In future work, we plan to analyze the trade-offs between searching and analyzing in more depth.
\bibliography{biblio}
\appendix
\end{document}